\newcommand{\PRE}[1]{{#1}} 
\documentclass[12pt,aps,prd,preprint,tightenlines,superscriptaddress,
amsmath,amssymb,
nofootinbib]{revtex4-1}

\usepackage{enumitem}

\RequirePackage[colorlinks=true
,urlcolor=blue
,anchorcolor=blue
,citecolor=blue
,filecolor=blue
,linkcolor=blue
,menucolor=blue
,pagecolor=blue
,linktocpage=true
,pdfproducer=medialab
,pdfa=true
]{hyperref}

\usepackage[dvipdfmx]{graphicx}
\usepackage{amsmath,amssymb,amsthm,amsfonts}
\allowdisplaybreaks
\usepackage{slashed} 
\usepackage{graphicx}
\usepackage{subfigure}
\usepackage{dcolumn}
\usepackage{hyperref}      
\usepackage{bm}
\usepackage{epsfig}
\usepackage{epstopdf}
\usepackage{setspace}
\usepackage[usenames, dvipsnames]{color}
\usepackage{slashed}
\usepackage{comment}

\newcommand{\be}{\begin{equation}\begin{aligned}}
\newcommand{\ee}{\end{aligned}\end{equation}}
\newcommand{\beq}{\begin{equation}}
\newcommand{\eeq}{\end{equation}}
\newcommand{\beqa}{\begin{eqnarray}}
\newcommand{\eeqa}{\end{eqnarray}}

\renewcommand{\eqref}[1]{eq.~(\ref{#1})}

\newcommand{\lsim}{\raise0.3ex\hbox{$\;<$\kern-0.75em\raise-1.1ex\hbox{$\sim\;$}}}
\newcommand{\gsim}{\raise0.3ex\hbox{$\;>$\kern-0.75em\raise-1.1ex\hbox{$\sim\;$}}}

\usepackage{ulem}  

\begin{document}

\begin{flushright}
KYUSHU-RCAPP-2020-03\\ 
UME-PP-014
\end{flushright}

\title{\PRE{\vspace*{1.0in}}
Dark Photon from Light Scalar Boson Decays at FASER
\PRE{\vspace*{.4in}}}

\author{Takeshi Araki}
\email{t-araki@den.ohu-u.ac.jp}
\affiliation{Faculty of Dentistry, Ohu University, 31-1 Sankakudo, Tomita-machi, Koriyama, Fukushima
963-8611,  Japan
\PRE{\vspace*{.1in}}}

\author{Kento Asai}
\email{asai@hep-th.phys.s.u-tokyo.ac.jp}
\affiliation{Department of Physics, University of Tokyo, 7-3-1 Hongo, Bunkyo-ku, Tokyo
 133--0033, Japan
\PRE{\vspace*{.1in}}}

\author{Hidetoshi Otono}
\email{otono@phys.kyushu-u.ac.jp}
\affiliation{Research Center for Advanced Particle Physics, Kyushu University, Fukuoka
819-0395, Japan
\PRE{\vspace*{.1in}}}

\author{Takashi Shimomura}
\email{shimomura@cc.miyazaki-u.ac.jp}
\affiliation{Faculty of Education, University of Miyazaki, 1-1 Gakuen-Kibanadai-Nishi, Miyazaki
889-2192, Japan
\PRE{\vspace*{.1in}}}

\author{Yosuke Takubo}
\email{yosuke.takubo@kek.jp}
\affiliation{Institute of Particle and Nuclear Study, KEK, Oho 1-1, Tsukuba, Ibaraki 
305-0801, Japan
\PRE{\vspace*{.4in}}}


\begin{abstract}
\PRE{\vspace*{.2in}} 
FASER is one of the promising experiments which search for long-lived particles beyond the Standard Model.
In this paper, we focus on dark photon associating with an additional U(1) gauge symmetry, and also a scalar boson breaking this U(1) gauge symmetry.
We study the sensitivity to the dark photon originated from U(1)-breaking scalar decays.
We find that a sizable number of dark photon signatures can be expected in wider parameter space than previous studies.
\end{abstract}


\maketitle

\section{Introduction}
\label{sec:introduction}

FASER (ForwArd Search ExpeRiment)~\cite{Feng:2017uoz,Ariga:2018pin,Ariga:2019ufm,Ariga:2018uku} is a new experiment to search for new light, weakly interacting, neutral particles, that are generated at proton-proton collision in the Large Hadron Collider (LHC) at the European Organization for Nuclear Research (CERN). 
The detector will be placed 480~m downstream from the ATLAS interaction point (IP). 
Utilizing a large cross-section of proton-proton inelastic interaction in the forward region, FASER can realize high sensitivity to such new particles even with a compact detector.
FASER will collect about 150~fb$^{-1}$ of data with 14~TeV proton-proton colliding energy in LHC Run~3 and be upgraded to FASER~2 to take 3~ab$^{-1}$ at High-Luminosity LHC.

A typical example of the new light, weakly interacting, neutral particles is a dark photon, and the discovery potential of the dark photon at FASER was studied in detail in refs.~\cite{Feng:2017uoz,Ariga:2019ufm} (the detectability of other particles was also studied, e.g., dark Higgs bosons~\cite{Feng:2017vli, Boiarska:2019vid}, axion-like particles~\cite{Feng:2018pew}, inflatons~\cite{Okada:2019opp},  heavy neutral leptons~\cite{Kling:2018wct}, and long-lived bosonic particles related with the so-called KOTO anomaly~\cite{Kling:2020mch}).
The dark photon is a new U(1) gauge boson interacting with the Standard Model (SM) particles only through kinetic mixing with the SM photons~\cite{Okun:1982xi,Galison:1983pa,Holdom:1985ag,Foot:1991kb,Babu:1997st,Boehm:2003hm,Pospelov:2008zw}.
The kinetic mixing is tightly constrained to be much small by collider searches (for details and references, see a review, e.g. ref.~\cite{Fabbrichesi:2020wbt}), and thus, dark photons can only weakly interact with the SM particles.
Nevertheless, thanks to the large cross-section in the forward region at LHC, a large number of dark photons can be produced via light meson decays and proton bremsstrahlung, providing us with an exciting opportunity to explore the unconstrained parameter space of dark photon models.

Dark photons can easily be incorporated into the SM.
One just has to add a new gauge field $A'$ to the SM particle content, and only two parameters are necessary to describe the minimal model, i.e., a kinetic mixing parameter $\epsilon$ and a dark photon mass $m_{A'}$.
On one hand, the minimal model is very simple and useful to find out the basic properties of dark photons.
On the other hand, however, it is also likely that there exist scalar bosons whose vacuum expectation values (VEVs) spontaneously break the new U(1) gauge symmetry; it gives us the dynamical origin of the dark photon mass.
In the latter case, the scalar bosons have a coupling with two dark photons as a consequence of the mass generation. 
Importantly, this coupling is not suppressed by the small kinetic mixing, and the scalar bosons can dominantly decay into a pair of the dark photons if kinematically allowed.
Moreover,  the scalar bosons can be produced by rare decays of hadrons through mixing with the SM Higgs, so that the sizable increase of the dark photon signals at FASER can be expected in this case. 

In this work, we consider a dark photon model with a new scalar boson that breaks the U(1) gauge symmetry, and investigate how the existence of the new scalar boson affects the dark photon search at FASER.
In the minimal case where dark photons are originated only from meson decays, the number of produced dark photons depends on the square of the kinetic mixing, and larger mixing leads to more production of dark photons. 
Conversely, such a larger mixing makes a lifetime of dark photons shorter, and it leads to a decrease in the number of dark photons reaching the FASER detector.
In contrast, in our model, dark photons can be produced through not only the kinetic mixing but also the gauge interaction.
We find two distinctive cases to the different scalesx of the gauge coupling constant. 
One case is that the scalar boson, with a short lifetime, dominantly decays into the dark photons, and the dark photon has a long lifetime 
to reach the detector.
The other case is that the scalar boson has a long lifetime and decays into the dark photons near or inside the detector, and the dark photon can be short-lived with large kinetic mixing.
Note that the inclusion of a new scalar boson to extend FASER's discovery potential of dark photons was also proposed in ref.~\cite{Jodlowski:2019ycu}. 
Although the model considered in our work is the same as that in ref.~\cite{Jodlowski:2019ycu}, we are interested in different parameter regions.
In contrast to ref.~\cite{Jodlowski:2019ycu} in which the scalar bosons are assumed to be long-lived and to produce dark photons by scattering with nuclei in front of the detector or inside it, we are here interested in the case where the scalar bosons decay into a pair of dark photons.
Consequently, we find that there is a parameter space where FASER can cover but has not been taken into account in the previous studies~\cite{Feng:2017uoz,Ariga:2019ufm,Jodlowski:2019ycu}.

This paper is organized as follows.
In section~\ref{sec:DPmodel}, we briefly review the dark photon model and show the decay widths of the dark photon and additional U(1)-breaking scalar boson.
In section~\ref{sec:detector}, we show the details of the FASER detector.
In section~\ref{sec:production}, the expected number of dark photon events from the scalar boson decays is discussed.
In section~\ref{sec:result}, we show the results of our analyses and, lastly, we summarize our discussion in section~\ref{sec:summary}.

\section{Dark photon model}
\label{sec:DPmodel}

We consider an extension of the SM, introducing an extra U(1) gauge symmetry under which all of the SM particles are uncharged. 
The extra gauge symmetry is spontaneously broken by the VEV of a complex scalar only charged under the extra U(1) symmetry. 
Then, the extra gauge boson acquires a mass. 
We refer to this massive gauge boson as the dark photon, $A'$, throughout this paper. 
One of the consequences of the spontaneous symmetry breaking is the interaction term between two dark photons and the CP-even component $\phi$ of the U(1)-breaking scalar. 
The interaction strength is determined by the dark photon mass, $m_{A'}$, and the extra U(1) gauge coupling, $g'$.
After the electroweak symmetry breaking, the dark photon can mix with the photon through the gauge kinetic mixing term between the SM hypercharge and the extra U(1) gauge fields.
Then, the dark photon can interact with the charged SM fermions via the electromagnetic current. 
Furthermore, the scalar, $\phi$, can mix with the SM Higgs boson through the quartic coupling of the scalar bosons, and hence the interaction terms of $\phi$ with the SM fermions, $f$, are induced. 
The relevant interaction Lagrangian to our study is given by
\begin{align}
\label{eq:Lagrangian}
   \mathcal{L}_{\rm int} = g' m_{A'} \phi A'_\mu A'^\mu 
         + \sum_f \frac{m_f \theta}{v} \phi \bar{f} f - \epsilon e A'_\mu J_{\rm EM}^\mu~, 
\end{align}
where $m_f$ are the masses of the SM fermions, and $v$ and $\theta$ are the VEV of the SM Higgs boson and scalar mixing parameter, respectively. 
In the last term, $\epsilon,~e$ and $J_{\rm EM}^\mu$ stand for the gauge kinetic mixing, electric charge of proton, and the electromagnetic current of the SM, respectively.

With \eqref{eq:Lagrangian}, the two-body decay widths of $\phi$ into a pair of $A'$ and  lighter SM fermions 
are given by
\begin{align}
\label{eq:decaywidth-scalar-Ap}
   \Gamma_{\phi \to A'A'} &= \frac{{g'}^2 }{8 \pi} \frac{m_{A'}^2}{m_\phi} \beta_\phi(A') 
        \left( 2 + \frac{m_\phi^4}{4 m_{A'}^4} \left( 1 - \frac{2 m_{A'}^2}{m_\phi^2} \right)^2 \right)~, \\
\label{eq:decaywidth-scalar-f}        
   \Gamma_{\phi \to f\bar{f}} &= \frac{m_\phi}{8 \pi} \left( \frac{m_f}{v} \right)^2 \theta^2 
        \left( 1 - \frac{4 m_f^2}{m_\phi^2} \right) \beta_\phi(f)~,
\end{align}
where $m_\phi$ is the mass of $\phi$, and $\beta_i(j) = \sqrt{1 - 4 m_j^2/m_i^2}$ is the kinematic factor of the decay $i \to jj$.
Here we assumed that $\phi$ is much lighter than the weak bosons as well as the SM Higgs boson. 
Thus, the decays into these particles are kinematically forbidden.
From \eqref{eq:decaywidth-scalar-Ap}, one finds the decay width of $\phi$ into $A'A'$ is significantly enhanced by the factor $m_\phi^2/m_{A'}^2$ when $m_\phi \gg m_{A'}$. 
On the contrary, the decay widths into the SM fermions, \eqref{eq:decaywidth-scalar-f}, are suppressed due to the scalar mixing $\theta^2$, which is constrained as $\theta \lesssim O(10^{-4})$ by LHCb~\cite{Aaij:2015tna,Aaij:2016qsm}. 
In this situation, $\phi$ dominantly decays into the dark photons.

\begin{figure}[tb]
\centering
\includegraphics[width=0.48\textwidth]{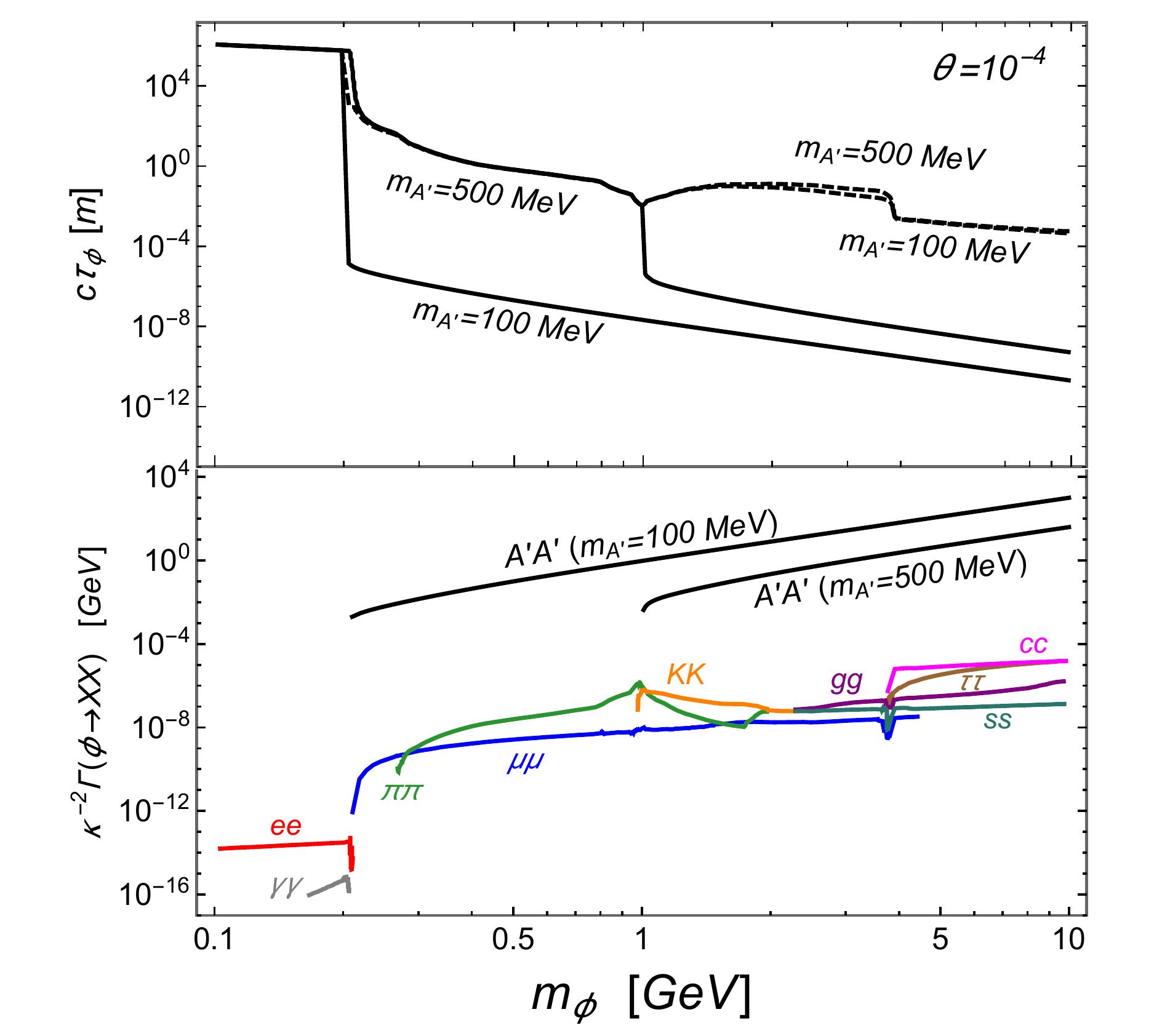} \quad 
\includegraphics[width=0.48\textwidth]{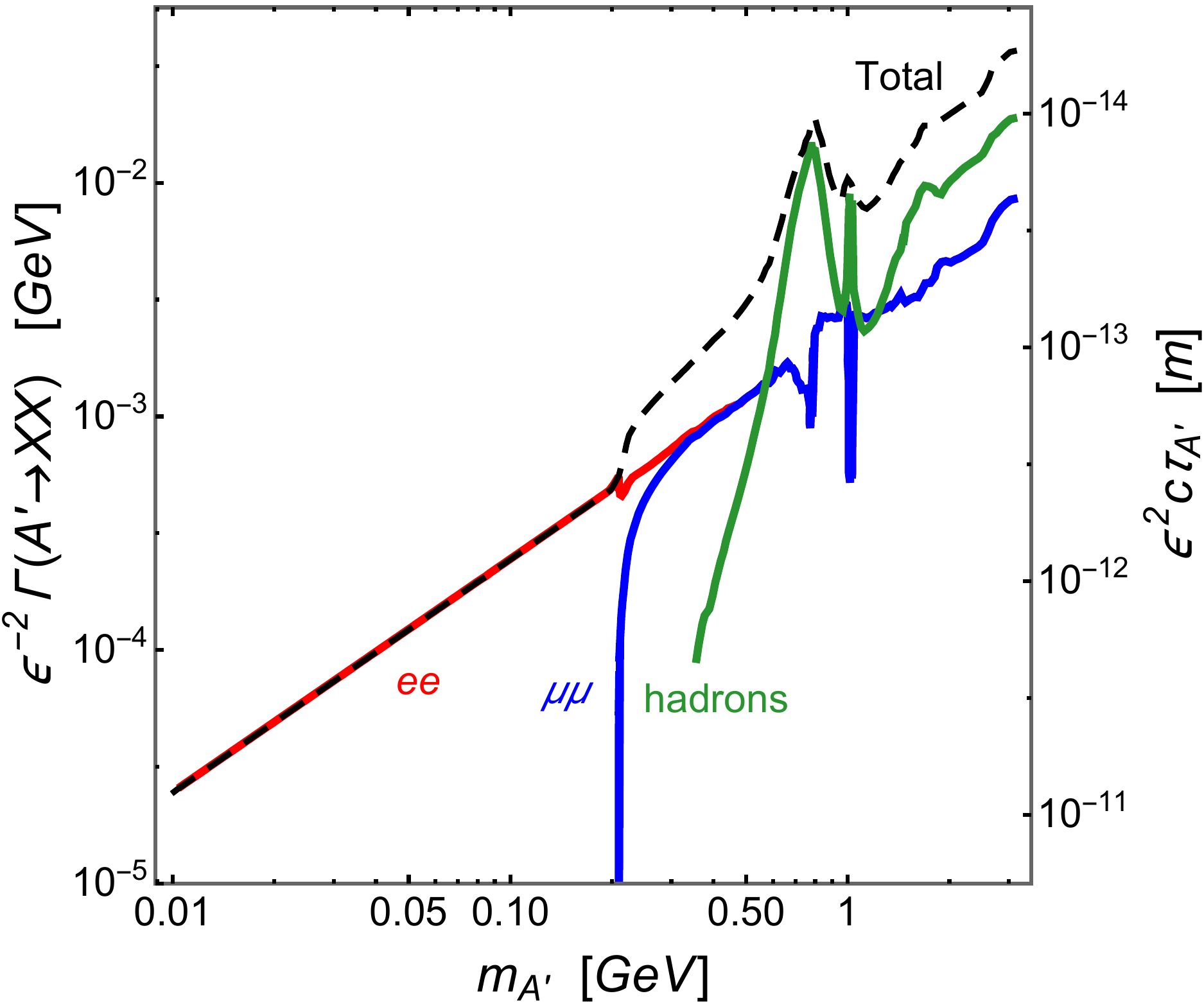} 
\caption{
Left: The decay length (top panel) and partial widths (bottom panel) of the scalar boson. In the left-upper panel, the black solid (dashed) curves are the decay length for $m_{A'} = 100$ and $500$ MeV for $g' = 10^{-4}~(10^{-8})$ in the rest frame of $\phi$. 
The scalar mixing is fixed to be $10^{-4}$ as an illustrating example.
In the left-bottom panel, the color curves are partial decay widths into the SM particles (indicated near each curve). 
The black solid curves are the partial widths into the dark photons with $m_{A'} = 100$ and $500$ MeV. 
The widths are normalized by the scalar mixing ($\kappa = \theta$) for the SM final states and by the gauge coupling constant ($\kappa = g'$) for $A'$.
Right: The decay widths of the dark photon into the SM particles normalized by $\epsilon^2$.
The black dashed curve is the total decay width. 
The right vertical axis is the decay length in the rest frame of $A'$ normalized by $\epsilon^{-2}$.
}
\label{fig:decays}
\end{figure}
The decay length and partial decay widths of $\phi$ are shown in the top and bottom panels of the left figure in figure~\ref{fig:decays}, following refs.~\cite{Bezrukov:2013fca,Winkler:2018qyg,Ariga:2018uku}, respectively. 
One can see in the bottom panel that the decay width into the dark photons is much larger than those into the SM particles.  
In the top panel, $g'$ is taken to be $10^{-4}$ (solid) and $10^{-8}$ (dashed) with $\theta=10^{-4}$. 
It is shown for $g' = 10^{-4}$, the partial decay width into $A'$ is so large that the decay length of the scalar is pretty short. 
Even in the laboratory frame where $\phi$ will have the energy of $\mathcal{O}(1)$ TeV, the scalar will decay within a few cm from the IP. 
Comparing the distance from the IP to the end of the FASER detector (480 m), we can safely assume the scalar decays at the IP. 
For $g'=10^{-8}$, the partial decay width into $A'$ is comparable to or smaller than those into the SM fermions, and the scalar can travel to the FASER detector before its decay.

The dark photon, on the other hand, can decay into the SM fermions, and the partial decay widths of $A'$ are given as follows:
\begin{align}
\label{eq:decaywidth-Ap-ff}
   \Gamma_{A' \to f\bar{f}} &= \frac{\epsilon^2 e^2}{12 \pi} m_{A'} \left( 1 + 2 \frac{m_f^2}{m_{A'}^2} \right) \beta_{A'}(f)~, \\
   \Gamma_{A' \to {\rm hadrons}} &=\Gamma_{A' \to \mu \bar{\mu}} R(s=m_{A'}^2)~,
\end{align}
where $R(s) \equiv \sigma(e^+ e^- \to {\rm hadrons})/\sigma(e^+ e^- \to \mu \bar{\mu})$, being $s$ the center of mass energy, 
i.e. $m_{A'}$ for the decay. 
The decay width (left axis) and length (right axis) of the dark photon are shown in the right panel of figure~\ref{fig:decays}, following refs.~\cite{Buschmann:2015awa,Bauer:2018onh,Ariga:2018uku}.
In the figure, the decay width (length) is normalized by $\epsilon^2~(\epsilon^{-2})$. 
In the laboratory frame where $A'$ will have $\mathcal{O}(100)$ GeV, it can survive until arriving at the FASER detecter when $\epsilon < 10^{-3}$.

\section{FASER Detector}
\label{sec:detector}

Dimension of the FASER detector~\cite{Ariga:2018pin} is 10~cm in radius and 5~m long. 
At the entrance to the detector, two scintillator stations are used to veto charged particles coming through the cavern wall from the IP, primarily high-energy muons, with 99.99\%  veto efficiency of charged particles per station.
Between the stations is a lead plate with 20 radiation length thickness that converts photons coming from upstream of the detector into electromagnetic showers that are vetoed by the scintillators. 

The veto stations are followed by a 1.5~m long, 0.55~T permanent dipole  magnet with a 10~cm aperture radius. 
This is the decay volume for new particles decaying into a pair of charged particles, where the magnetic field separates the decay products to a detectable distance. 

After the decay volume is a spectrometer consisting of two 1~m long 0.55~T dipole magnets with three tracking stations, that are located at either end and in between the magnets. 
Each tracking station is composed of three layers of precision silicon strip detectors. 
Scintillator stations for triggering and precision time measurements are located at the entrance and exit of the spectrometer. 
The primary purpose of the spectrometer is to observe the characteristic signal of two oppositely-charged particles pointing back towards the IP, measure their momentum, and sweep out low-momentum charged particles before they reach the final layer of the spectrometer.

The final component is the electromagnetic calorimeter. 
This will identify high-energy electrons and photons and measure the total electromagnetic energy. 

FASER is being installed in the unused service tunnel TI12 from autumn 2020, which is 480~m downstream from the ATLAS IP. 
The goal of data-taking is to collect about 150~fb$^{-1}$ of data with 14~TeV proton-proton colliding energy during 2022-2024 in LHC Run~3. 

Upgrade of the FASER detector (FASER~2) is also planned to extend sensitivity to new particles at the High-Luminosity LHC (HL-LHC). 
The detector radius will be enlarged to 1~m, so that the acceptance for the new particles will become hundred times larger than that of FASER. 
In addition, the decay volume will be extended to 5 m long.
In the operation at HL-LHC, FASER~2 aims to collect 3~ab$^{-1}$ of data with 14 TeV, about 20 times bigger amount of data that will be taken in FASER. 
The dimensions and integrated luminosities of the FASER detector are summarized in table~\ref{tab:faser-dimension}. 
\begin{table}[t]
\begin{tabular}{|c|c|c|c|c|} \hline
 \hspace{2cm} & ~~~$L_{\rm min}$~(m)~~~ & ~~~$L_{\rm max}$~(m)~~~ & ~~~$R$~(m)~~~ & ~~~$\mathcal{L}$~(ab$^{-1}$)~~~ \\ \hline \hline 
 FASER  & 478.5 & 480 & 0.1 & 0.15 \\ \hline
 FASER~2 & 475 & 480 & 1.0 & 3.0 \\ \hline
\end{tabular}
\caption{
Dimension of the FASER detector and integrated luminosity used for this study. 
$L_{\rm min}$ and $L_{\rm max}$ are the distance to the front and rear end of the FASER detector from the IP, respectively. 
$R$ is detector radius, respectively. 
$\mathcal{L}$ is the integrated luminosity.
}
\label{tab:faser-dimension}
\end{table}

\section{Dark Photon Production and Detection}
\label{sec:production}

In this section, we discuss the production of the dark photons from the scalar boson decays.
The scalar boson is dominantly produced by the decays of mesons through the SM Higgs-$\phi$ mixing.
When the scalar mixing $\theta$ is much smaller than the unity, the decay branching ratios of $\phi$ production are given by \footnote{There also exists the production from $\eta$ meson decays. However, it is less effective and we omitted that decay.  }~\cite{Feng:2017vli}
\begin{align}
\label{eq:br-sigma-prod-B}
   {\rm Br}(B \to X_s \phi) &\simeq 5.7 \left( 1 - \frac{m_\phi^2}{m_b^2} \right)^2 \theta^2~, \\
   {\rm Br}(K^\pm \to \pi^\pm \phi) &= 2.0 \times 10^{-3} \frac{2 p_\phi^0}{m_K} \theta^2~, \\
   {\rm Br}(K_L \to \pi^0 \phi) &= 7.0 \times 10^{-3} \frac{2 p_\phi^0}{m_K} \theta^2~, \\
   {\rm Br}(K_S \to \pi^0 \phi) &= 2.2 \times 10^{-6} \frac{2 p_\phi^0}{m_K} \theta^2~,  \\
\label{eq:br-sigma-prod-etap}
   {\rm Br}(\eta' \to \eta \phi) &= 7.2 \times 10^{-5} \frac{2 p_\phi^0}{m_{\eta'}} \theta^2~, 
\end{align}
where $p_\phi^0 = \lambda^{1/2}(m_{K(\eta')}^2, m^2_{\pi(\eta)}, m_\phi^2) / (2 m_{K(\eta')})$ for the $K (\eta')$ meson decay, with $\lambda(a, b, c) = a^2 + b^2 + c^2 - 2 (ab + bc + ca)$, is the three-momentum of the scalar in the parent meson's rest frame.
The produced scalars travel to the direction of the FASER detector and decay into a pair of dark photons.
The probability that the dark photon decays inside the FASER detector is given by 
\begin{align}
\label{eq:prob}
\mathcal{P}_{A'}^{\rm det}(\bm{p}_{A'}, \bm{p}_\phi) &= 
\frac{1}{\bar{d}_\phi \cos\theta_\phi} \int_{z_{\phi,\mathrm{min}}}^{z_{\phi,\mathrm{max}}} dz_\phi e^{-\frac{z_\phi}{\bar{d}_\phi \cos\theta_\phi}}
\frac{1}{\bar{d}_{A'} \cos\theta_{A'}} \int_{z_{A',\mathrm{min}}}^{L_{\mathrm{max}}} dz_{A'} e^{-\frac{z_{A'}-z_\phi}{\bar{d}_{A'} \cos\theta_{A'}}} 
\nonumber \\
&\quad \times \Theta(R - r_{A',R}) \Theta(R - r_{A',F} ),
\end{align}
where $\bm{p}_i$ and $\theta_i$ $(i=A',\phi)$ denote the momentum and angle with respect to 
the beam axis ($z$-axis), respectively, and $\bar{d}_i$ denotes the decay length of the dark photon or scalar in the laboratory frame. The absolute values of the momenta are determined from the energy-momentum conservations, and 
the angular distributions are generated by Monte Carlo simulations with $1000$ samples.
The detector radius and the distance between the IP and rear (front) end of the FASER detector are denoted as 
$R$ and $L_{\rm max (min)}$, respectively, which are shown in table~\ref{tab:faser-dimension}.
In the step functions, $r_{A',R}$ and $r_{A',F}$ represent the distance of the dark photon from the beam axis at $z_{A'} = L_{\mathrm{max}}$ and $L_{\mathrm{min}}$, respectively. 
The integral regions can be determined so that the step functions are satisfied 
for given $\bm{p}_{A'}$ and $\bm{p}_\phi$. 
Denoting the position of $i$ as $(x_{i}, y_{i},z_{i})$, $(x_{i}, y_{i})$ is given as ($z_{i} p_{i,x}/p_{i,z}$, $z_{i} p_{i,y}/p_{i,z}$).
Then, $r_{A' ,R(F)}= \sqrt{x_{A'}^2 + y_{A'}^2}$ is obtained at $z_{A'} = L_{\mathrm{max(min)}}$. When $z_{\phi,\mathrm{min}}$ 
is negative, the lower limit of the $z_{\phi}$ integration is set to zero.

Given this probability, the total number of events of dark photon decays inside the FASER detector is given by
\begin{align}
\label{eq:num-of-event}
   N &= \mathcal{L} \int d\bm{p}_{A'} \frac{d\sigma_{pp \to A' X}}{dp_{A'} d\theta_{A'}} 
   \mathcal{P}_{A'}^{\rm det}(\bm{p}_{A'}, \bm{p}_\phi) \nonumber \\
   &= \mathcal{L} \sum_{i:{\rm meson}} \sum_{j=1,2} \int dp_i d\theta_i \int d\bm{p}_{A'} \int d\bm{p}_\phi
   \frac{d\sigma_{pp \to i X}}{dp_i d\theta_i } {\rm Br}(i \to \tilde{X} \phi) {\rm Br}(\phi \to A_1' A_2') \nonumber \\   
   &\qquad \times \mathcal{P}_{A'_j}^{\rm det}(\bm{p}_{A'}, \bm{p}_\phi)~,
\end{align}
where $X$ and $\tilde{X}$ are mesons shown in eqs.~(\ref{eq:br-sigma-prod-B})-(\ref{eq:br-sigma-prod-etap}), and $\mathcal{L}$ is the expected integrated luminosity. 

Lastly, we comment on the efficiency of the FASER detector.
A silicon strip module has more than 99\% of detection efficiency~\cite{Campabadal:2005cn}. 
Considering about 9 silicon strip layers in the spectrometer, the detection efficiency can be assumed as 100\% for the signal events.  
Although natural rock and LHC shielding can eliminate most of the potential backgrounds, there still remain high energetic muons with radiation and neutrinos as the main backgrounds. 
In the simulation, 80k muon events with $\gamma$, electro-magnetic or hadronic shower as well as a few neutrino events with charged current or neutral current interaction are expected to enter the FASER detector from the direction of the IP with the energy of secondary particles above 100~GeV in 150~fb$^{-1}$~\cite{Ariga:2018zuc}. 
Assuming 99.99\% veto efficiency of each scintillator station, these backgrounds can be reduced to a negligible level.

\section{Result} \label{sec:result}

We here calculate the number of the U(1)-breaking scalar bosons from meson decays and, subsequently, the expected numbers of the signal events of the dark photon decaying into $ee,~\mu\mu$, and hadrons which can be detected at FASER and FASER~2.
In the calculation, we use the data sets of meson momenta, angles to the beam axis, and differential cross sections~\cite{Kling2019}, which are generated by the Monte Carlo event generator EPOS-LHC~\cite{Pierog_2015} implemented in the CRMC simulation package~\cite{CRMC} for light mesons such as kaons, while Pythia 8~\cite{Sj_strand_2006,Sj_strand_2008} with the Monash tune~\cite{Skands_2014} for $B$ mesons.
Moreover, the cut is imposed on the dark photon momentum, $p_{A'} > 100$~GeV, and the dark photons satisfying this condition effectively propagate to the FASER detector.

\begin{figure}[t]
\centering
\includegraphics[width=0.48\textwidth]{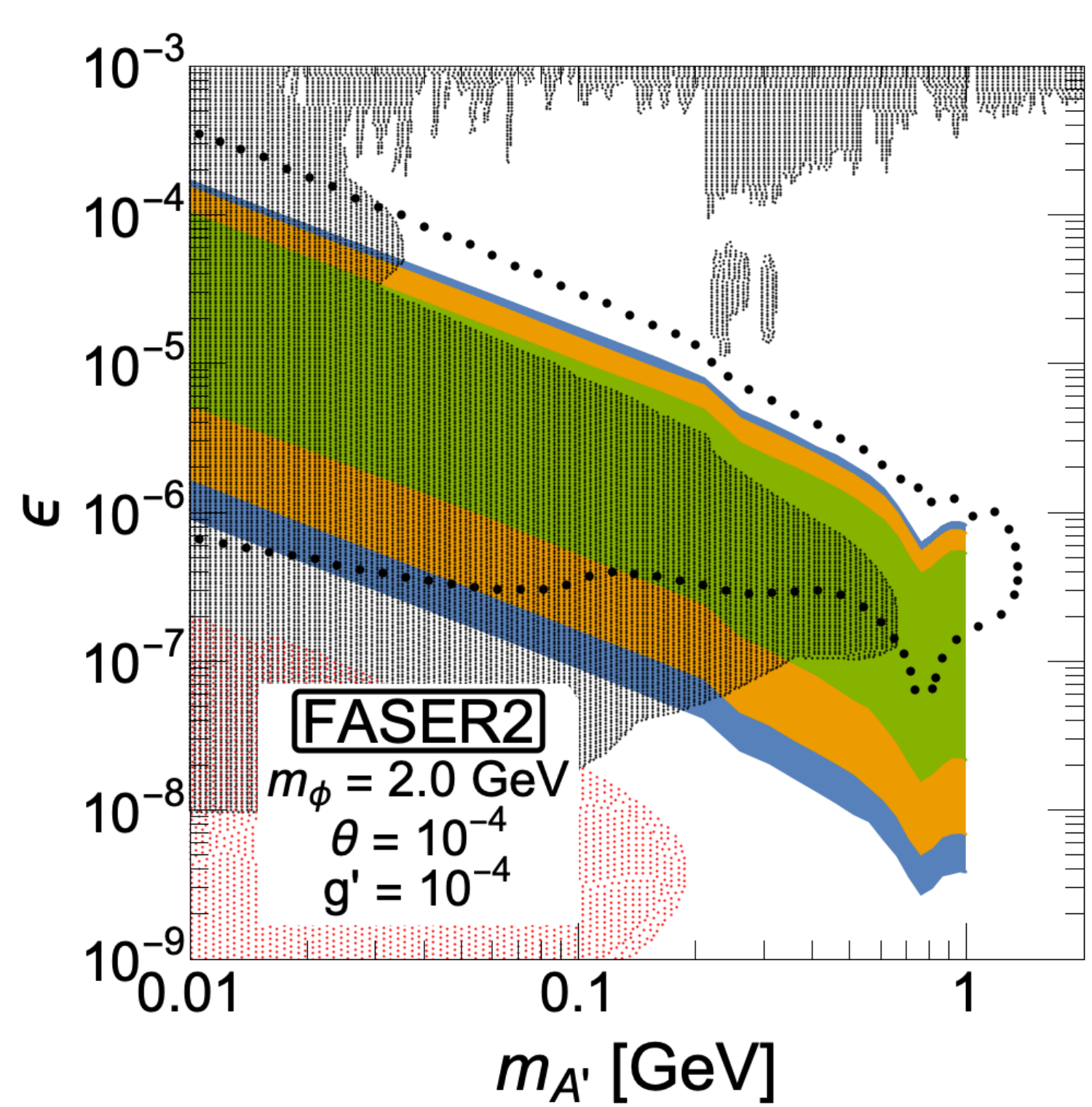} \quad
\includegraphics[width=0.48\textwidth]{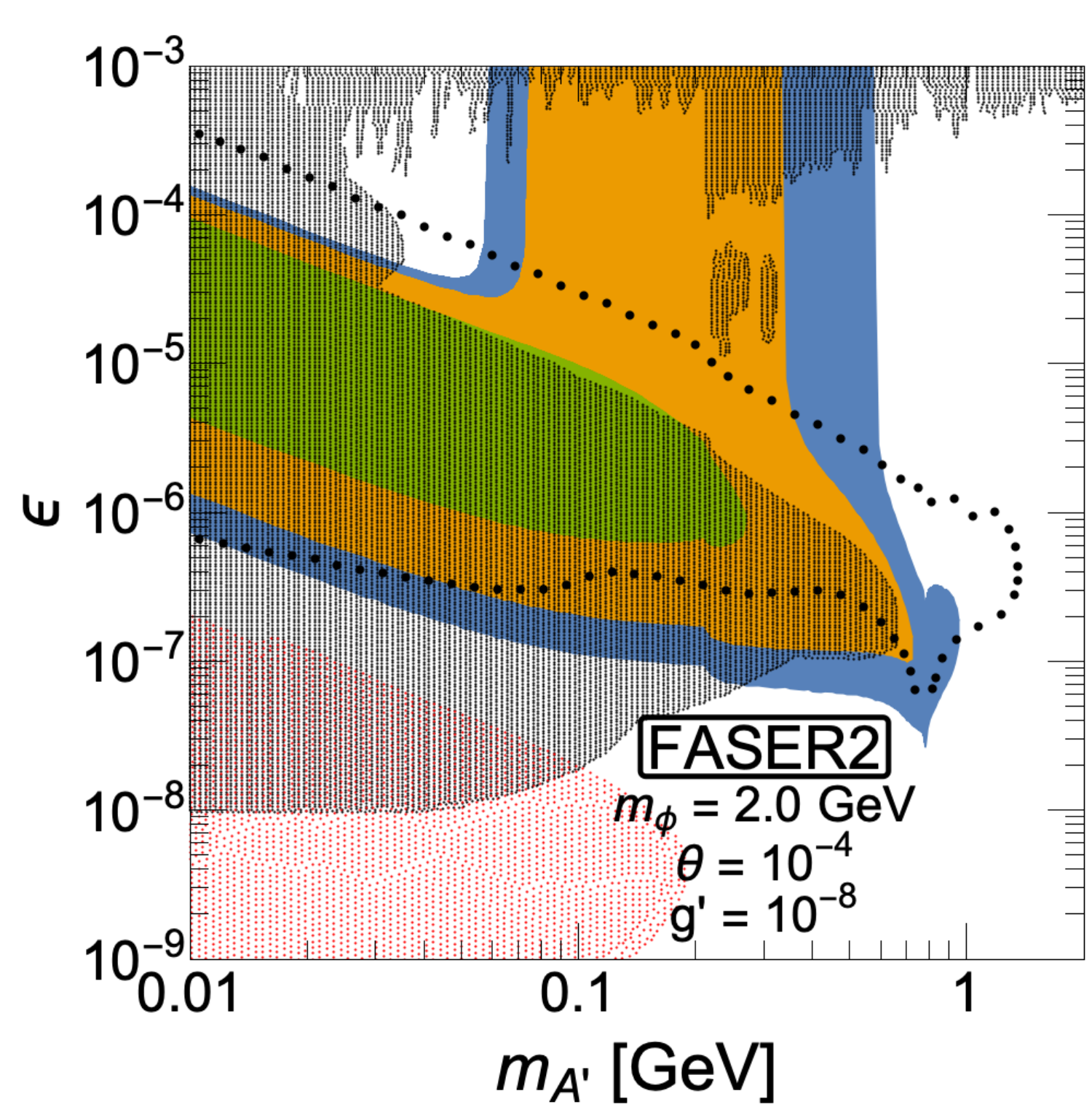} 
\caption{
Expected number of the signal events of the dark photon from the scalar boson decays in FASER~2.
Blue, orange, and green regions correspond to 3-9, 10-99, and $\geq 100$ events, respectively. 
Gray and pink shaded regions show the parameter spaces which are excluded by the current experiments~\cite{Fabbrichesi:2020wbt} and supernova cooling~\cite{Chang:2016ntp}, respectively.
Black dots show the projected dark photon sensitivity reaches at $95$\% C.L. in refs.~\cite{Ariga:2019ufm,Ariga:2018uku}.
}
\label{fig:result-faser2-heavy}
\end{figure}

We firstly show our numerical results of the expected number of the signal events at FASER 2 
in figure~\ref{fig:result-faser2-heavy}. 
As illustrating examples, we took the gauge coupling constant to be $10^{-4}$ (left panel) and $10^{-8}$ (right panel). 
The other parameters are fixed as $m_\phi = 2~{\rm GeV}$ and $\theta = 10^{-4}$ in both figures. 
For this mass, we only consider the scalar boson production from $B$ meson decays.
We found that FASER is not sensitive to these parameters.
In the calculation, to show the effects of the scalar boson decays, we do not include the dark photons produced by meson decays, bremsstrahlung, and QCD processes.
The blue, orange, and green regions correspond to the expected signal events for 3-9, 10-99, and $\geq 100$, respectively. 
The gray shaded region shows the parameter space excluded by the current experiments
\footnote{
Including the scalar production and its decays studied here might alter the bounds from the CHARM and $\nu$-Cal experiments.
In CHARM, $\mathcal{O}(10^4)$ of the scalar bosons are expected to be produced at most, which could results in $\mathcal{O}(10)$ signal events. 
In $\nu$-Cal, Bremsstrahlung productions from nucleons would provide $\mathcal{O}(1)$ signal events for $m_\phi = 2$ GeV. 
However, these numbers will be reduced by taking acceptances and efficiencies of the detectors into account. 
Thus, these effects would not be significant and the bounds for our study might change slightly~\cite{Kling-Trojanowski2020}.
},
including BaBar~\cite{Lees:2014xha}, NA48/2~\cite{Batley:2015lha}, NA64(e)~\cite{Bernhard:2020vca}, KLOE~\cite{Anastasi:2016ktq}, LHCb~\cite{Aaij:2019bvg}, E141~\cite{Riordan:1987aw}, E137~\cite{Marsicano:2018krp}, $\nu$-Cal~\cite{Blumlein:2013cua}, and the pink region is excluded by supernova cooling~\cite{Chang:2016ntp}.
For the details, see ref.~\cite{Fabbrichesi:2020wbt} and references therein.
The black dots represent the projected dark photon sensitivity reaches at $95$\% C.L. (corresponding to $3$ events) only from meson decays, bremsstrahlung, and QCD processes, shown in refs.~\cite{Ariga:2019ufm,Ariga:2018uku}.
In both panels, one can see that the number of dark photons produced from the scalar boson decays is much larger than those from meson decays, bremsstrahlung, and QCD processes in refs.~\cite{Ariga:2019ufm,Ariga:2018uku}. 
Furthermore, the sensitivity to small $\epsilon$ is improved in the high mass region. 
This is because the dark photon produced from the scalar boson decays can be as heavy as $m_\phi/2$, which are independent of $\epsilon$.
In the left panel, the kinetic mixing parameter is restricted by the reason above. 
As we explained in section~\ref{sec:DPmodel}, for $g'=10^{-4}$, the scalar boson instantly decays into a pair of dark photons.  
Then, the dark photons should be long-lived so that they can reach the FASER detector. 
Most of the dark photons with large $\epsilon$ decay before arriving at the detector.
The kinetic mixing parameter is also restricted due to the following reason.
The minimum energy of the dark photon from the scalar boson decay is $1$ GeV. 
The kinetic mixing parameter required with this energy can be roughly estimated from the right panel of figure~\ref{fig:decays}. 
One finds that it must be larger than $10^{-7}$ for $m_{A'} = 0.1$ GeV and $10^{-8}$ for $1$ GeV, respectively, which is in agreement with the results.
%

\begin{figure}[tb]
\centering
\includegraphics[width=0.48\textwidth]{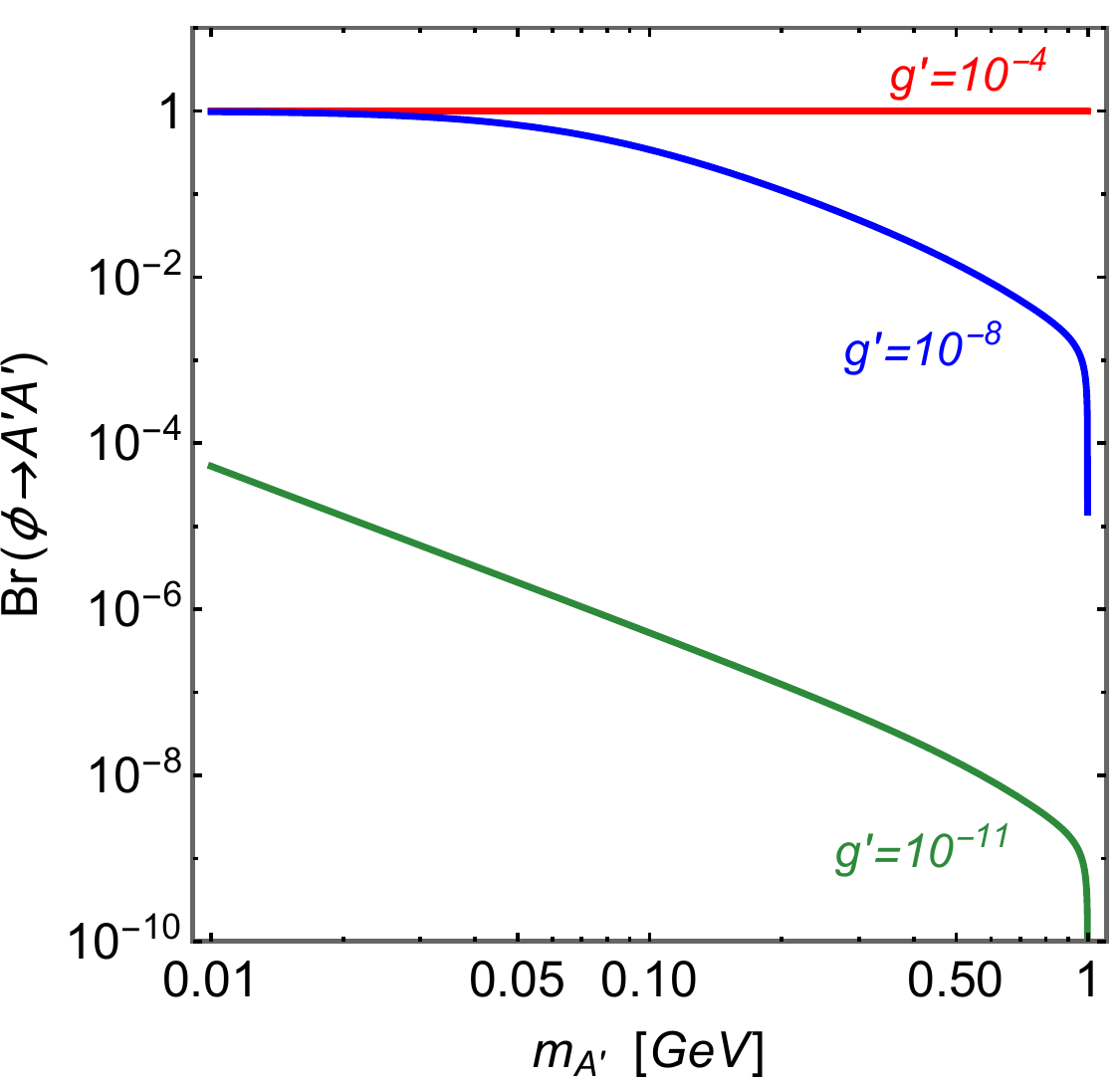} \quad
\includegraphics[width=0.48\textwidth]{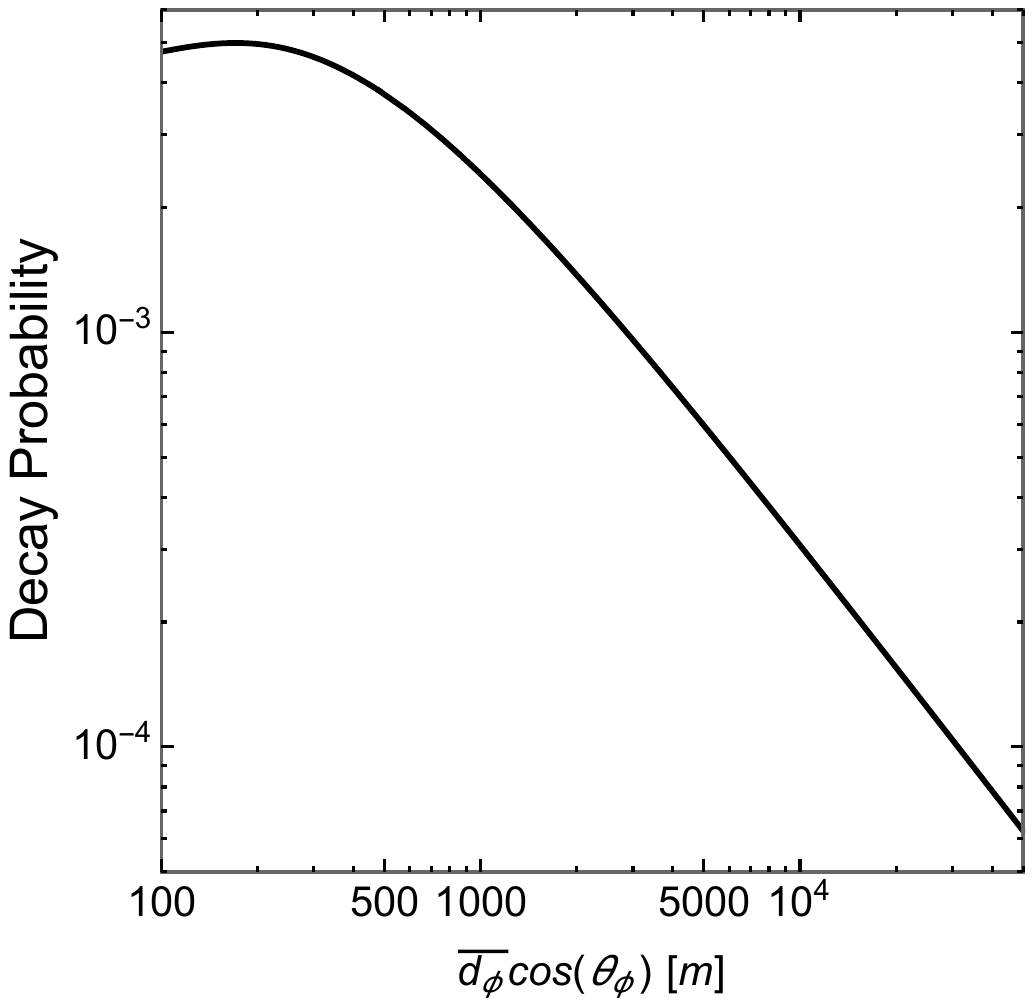} 
\caption{
Left: The branching ratio of $\phi \to A'A'$ as a function of $m_{A'}$ for $m_\phi = 2$ GeV and $\theta = 10^{-4}$. 
The red, blue, and green curves correspond to $g' = 10^{-4},~10^{-8}$ and $10^{-11}$, respectively. 
Right: The decay probability, \eqref{eq:prob}, as a function of the decay length $\bar{d}_\phi \cos\theta_\phi$ with $g'=10^{-8}$ at FASER~2. 
}
\label{fig:result2}
\end{figure}
The result of the left panel in figure~\ref{fig:result-faser2-heavy} remains almost the same for $g' \gsim 10^{-6}$ where the scalar boson decays at very close to the IP.
The situation, however, drastically changes when $g'$ is smaller than $10^{-6}$.
For such a small $g'$, the scalar boson is so long-lived that it can travel to the FASER detector and decay into the dark photons near or inside the detector, as shown in the left panel of figure~\ref{fig:decays}. 
Then, the dark photon can be short-lived with large kinetic mixing.
The right panel of figure~\ref{fig:result-faser2-heavy} shows the expected number of the signals for $g'=10^{-8}$ at FASER~2. 
One can see that the sensitivity region extends to a larger $\epsilon$ region, in which the decay length of the dark photon is much shorter than $L_{\rm min}$. 
In the left panel of figure~\ref{fig:result2}, the decay branching ratio of the scalar boson is shown with respect to $m_{A'}$ for $g'=10^{-4},~10^{-8}$, and $10^{-11}$. 
For $g'=10^{-8}$, the scalar boson mainly decays into the dark photons for $m_{A'} \lsim 0.2$ GeV. 
Thus, the expected number of the events decreases for $m_{A'} \gsim 0.2$ GeV. 
One can also see that there remains the sensitivity region in a small $\epsilon$ region, in which the dark photon is enough long-lived to reach the detector. 
Even in the long-lived case, some portion of the dark photon still can decay inside the detector. 
As an example, we showed the decay probability \eqref{eq:prob} as a function of  $\bar{d}_\phi \cos\theta_\phi$ for $z_{\phi,\mathrm{min}}  =0$ and $z_{\phi,\mathrm{max}} = L_{\mathrm{max}}$ assuming $\bar{d}_{A'} \cos \theta_{A'} = 480$ m in the right panel of figure~\ref{fig:result2}.
We see that the decay probability decreases by a factor of $1/7$ comparing $\bar{d}_\phi \cos\theta_\phi = 5000$ m with $500$ m. 
In the end, when $g' \lsim 10^{-11}$, the branching ratio of $\phi \to A'A'$ is too small to produce enough number of the dark photons, and hence the sensitivity region coincides with that of the direct dark photon productions from light meson decays.

\begin{figure}[t]
\centering
\includegraphics[width=0.48\textwidth]{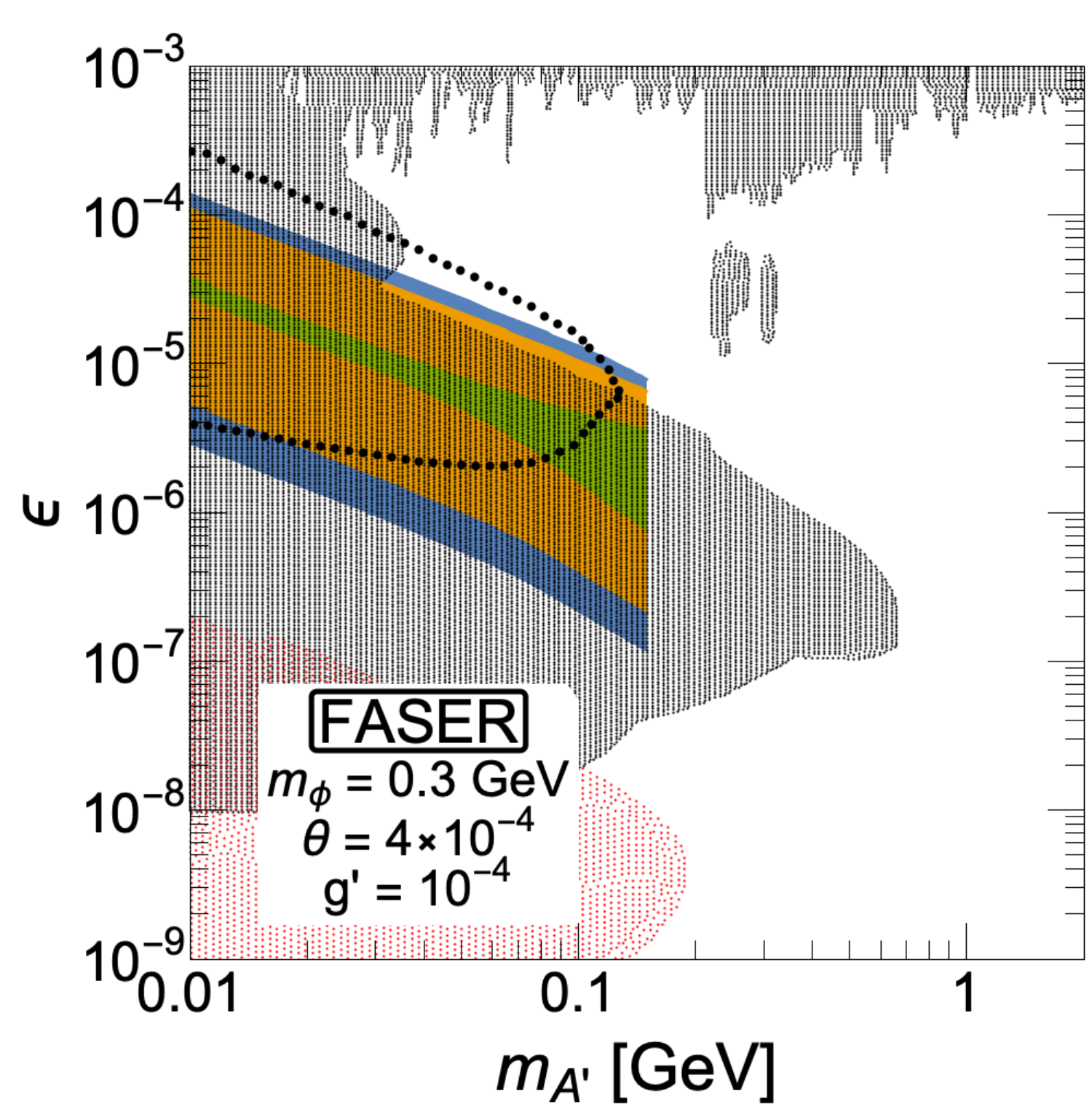} \quad
\includegraphics[width=0.48\textwidth]{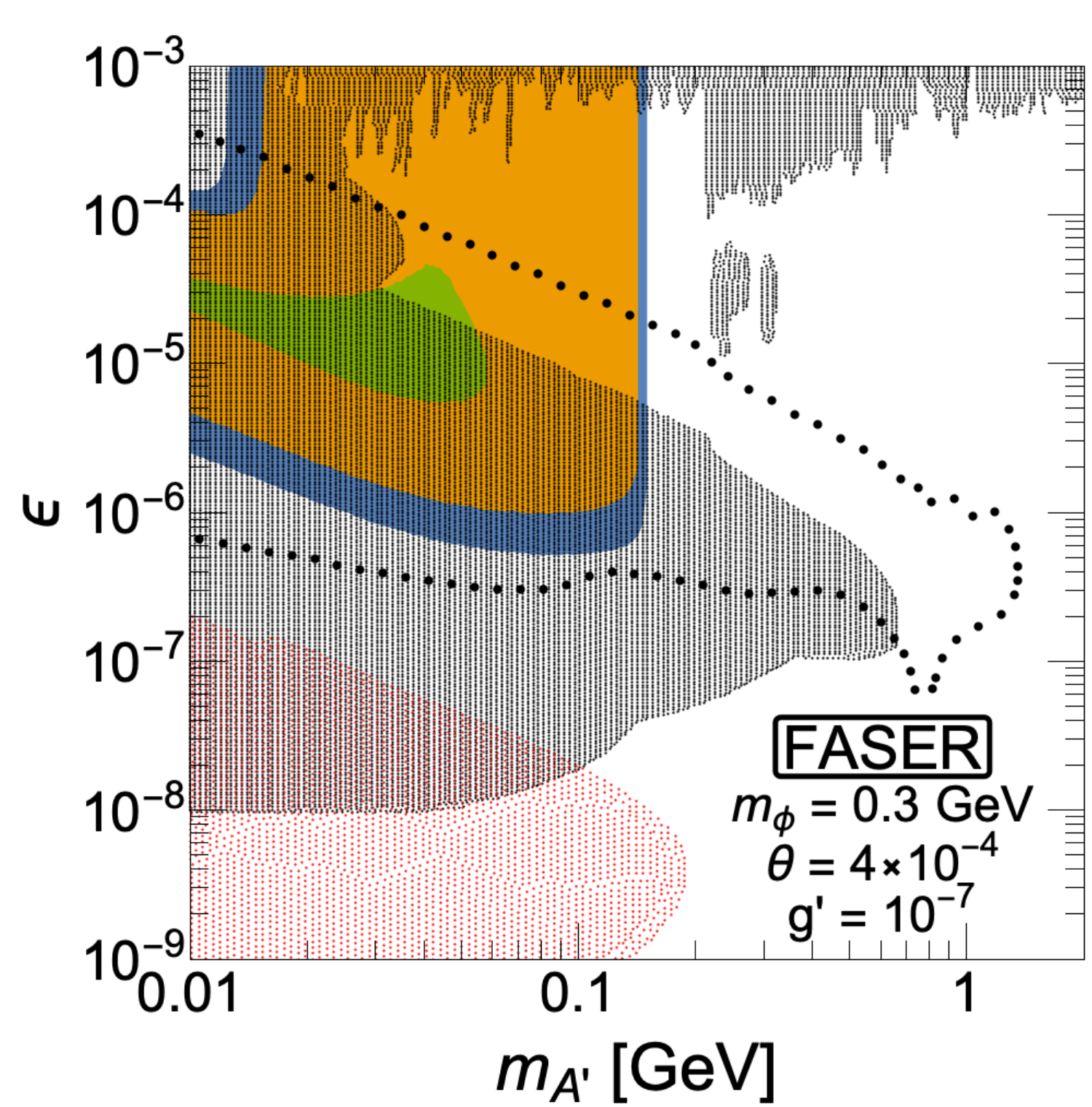} 
\caption{
The same plots as figure~\ref{fig:result-faser2-heavy} for FASER. The parameters are taken as $g'=10^{-4}$ (left) and 
$10^{-7}$ (right) while other parameters are fixed as $m_\phi = 0.3$ GeV and $\theta = 4 \times 10^{-4}$.
}
\label{fig:result-faser-light}
\end{figure}

\begin{figure}[t]
\centering
\includegraphics[width=0.48\textwidth]{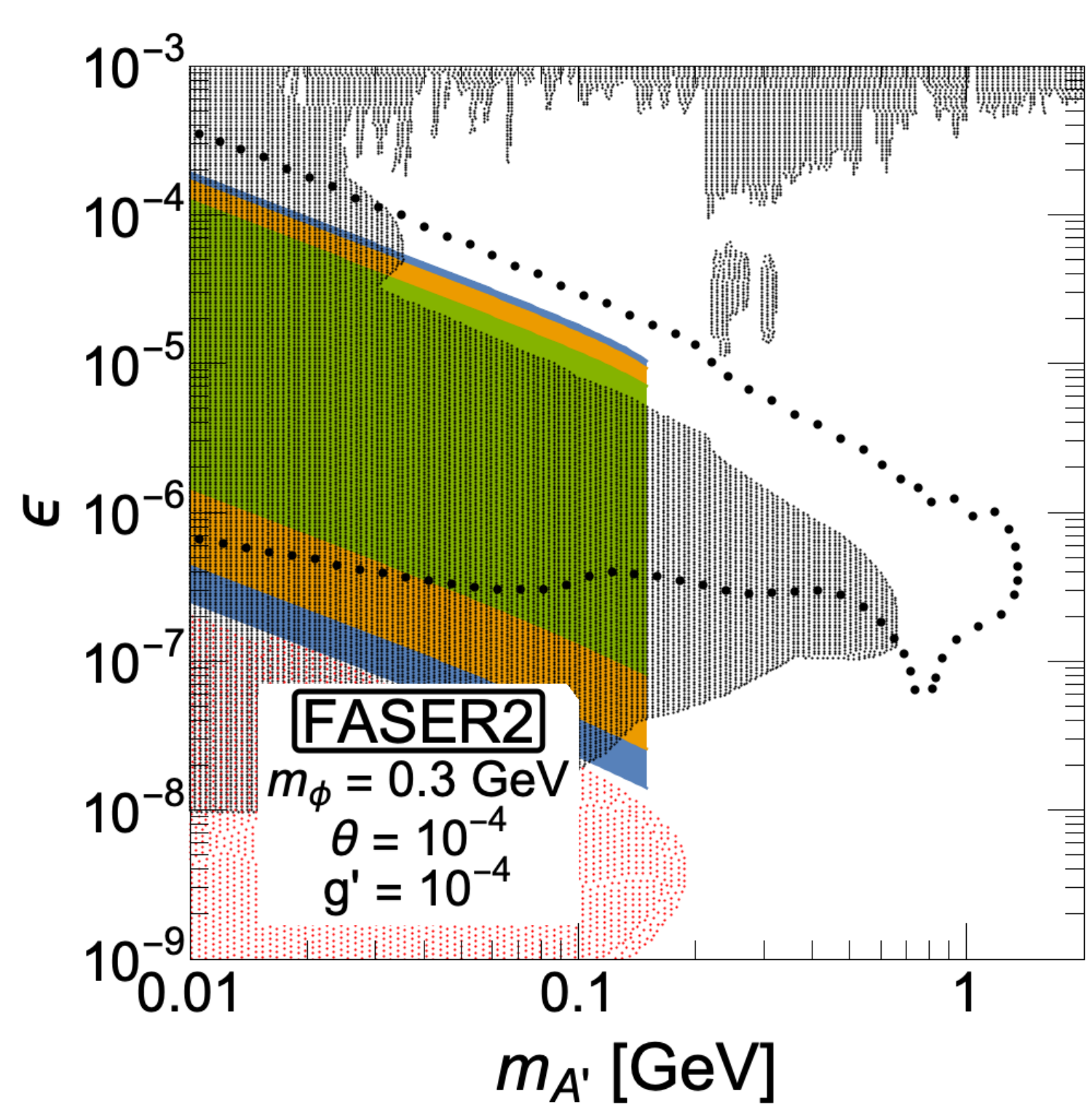} \quad
\includegraphics[width=0.48\textwidth]{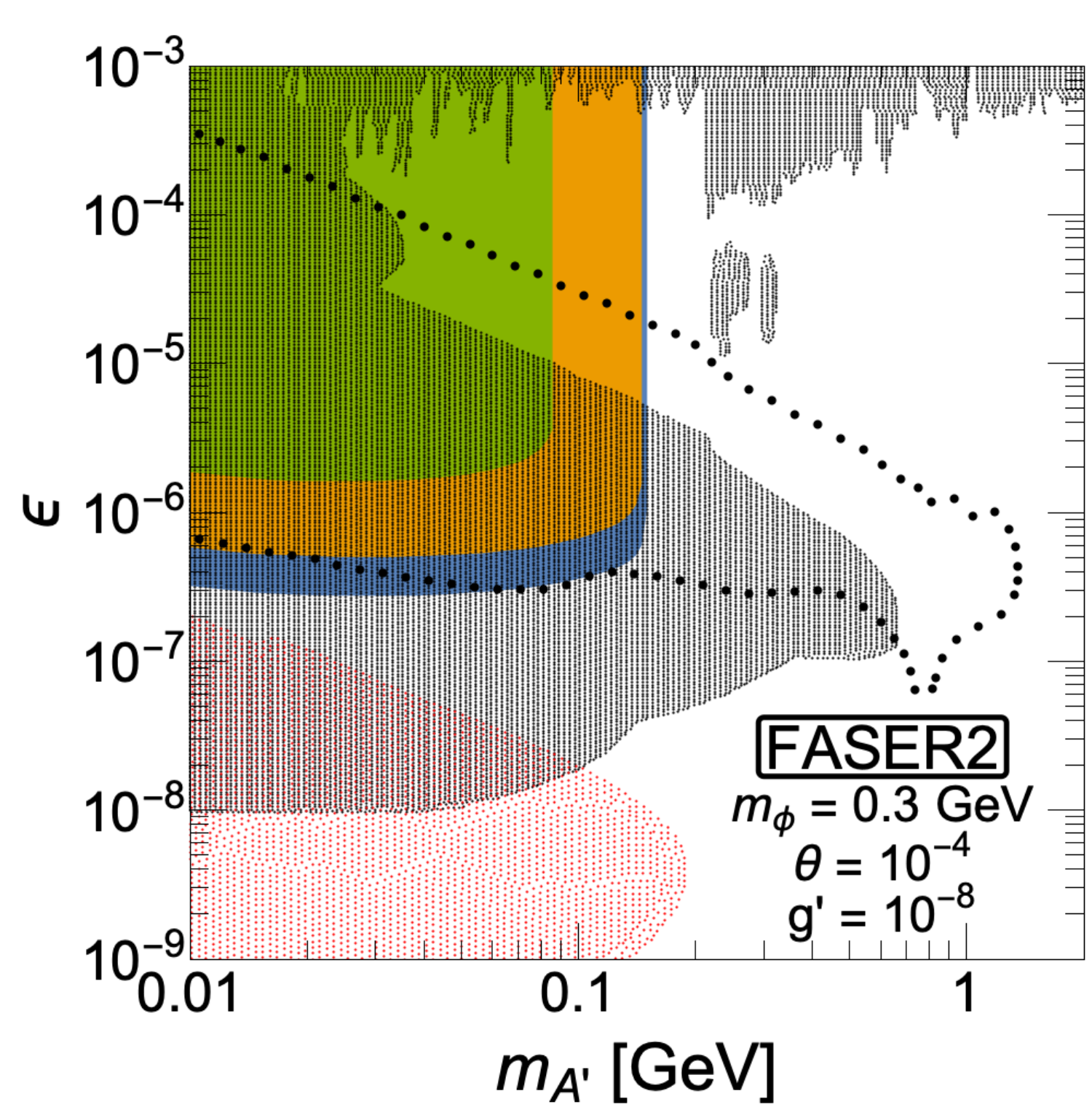} 
\caption{
The same plots as figure~\ref{fig:result-faser2-heavy} for FASER~2. The parameters are taken as $g'=10^{-4}$ (left) and 
$10^{-8}$ (right) while other parameters are fixed as $m_\phi = 2.0$ GeV and $\theta = 10^{-4}$.
}
\label{fig:result-faser2-light}
\end{figure}

Figures \ref{fig:result-faser-light} and \ref{fig:result-faser2-light} show the expected number of
 the signal events at FASER and FASER 2, respectively, for the lighter scalar boson with $m_\phi = 0.3$ GeV. 
For this mass, the production of the scalar boson is enhanced due to the decays of $K$ and $\eta'$ mesons
as well as 
$B$ mesons, and hence the number of the dark photons from the scalar boson decay is increased. 
The parameters are taken as $g' =10^{-4}$ (left) and $g'=10^{-7}$ (right) with $\theta = 4 \times 10^{-4}$ 
in figure~\ref{fig:result-faser-light}, and $g' =10^{-4}$ (left) and $g'=10^{-8}$ (right) with $\theta = 10^{-4}$ 
in figure~\ref{fig:result-faser2-light}.
One can see in the left panel of figure~\ref{fig:result-faser-light} that 
the scalar boson decay can provide $\mathcal{O}(1)$ signal events of the dark photon decays for 
just above the exclusion region 
at FASER. In FASER2, the left panel of figure~\ref{fig:result-faser2-light}, the scalar boson decays can provide $\mathcal{O}$(1-100) signal events  for $10^{-5} \leq \epsilon \leq 10^{-4}$ and $\mathcal{O}$(1-10) for $\epsilon \sim 2 \times 10^{-8}$. 
For the long-lived scalar case, the right panels of figures~\ref{fig:result-faser-light} and \ref{fig:result-faser2-light} show the sensitivity region also extends to a larger $\epsilon$ region at FASER and FASER2.
From these results, FASER is sensitive to the long-lived scalar boson with the light mass.
On the other hand, combined with the results of figure~\ref{fig:result-faser2-heavy}, FASER~2 is necessary to explore the short-lived scalar boson.

\section{Summary}
\label{sec:summary}
We have discussed the sensitivity of FASER to the dark photon from the scalar boson decays, assuming that the scalar spontaneously breaks the U(1) gauge symmetry associating the dark photon. 
The scalar can be produced in the decays of a meson through the mixing with the SM Higgs. 
Then, when the dark photon is much lighter than the scalar boson, the scalar dominantly decays into two dark photons due to the enhancement. 
A large number of dark photons can be produced from such $\phi$ decays. 

We have analyzed the expected number of events of $A'$ decays at FASER and FASER~2. 
Taking the parameters  $g' = 10^{-4}, ~\theta = 10^{-4}$ and $m_\phi = 2$ GeV as an illustrating example, we found the sensitivity is improved to smaller $\epsilon$ and  larger $m_{A'}$ regions than that in the case only with $\pi^0$ and $\eta$ meson decays.  
The decay of $\phi$ produces $\mathcal{O}(10^3)$ events of the dark photon decays at FASER~2. 
The large number of $\phi$ improves the sensitivity to $\epsilon$ up to $10^{-9}$. 
Furthermore, since $\phi$ is mainly produced from $B$ meson decays, it can be heavier than that from $\pi^0$ and $\eta$ mesons. 
For that reason, the heavier dark photon can be produced, which improves the sensitivity to $m_{A'}$. 
With our sample parameters, $m_{A'}$ can be explored up to $1$ GeV for $m_\phi = 2$ GeV. 
We also took $g' = 10^{-8}$ as another illustrating example. 
With this coupling, the scalar boson becomes long-lived and can reach the detector. 
Then, the dark photon signals can be detected even for large kinetic mixing. 
Moreover, we showed the expected number of the signals from the light scalar boson decays. 
In this case, the scalar boson can be produced from the decays of $K$ and $\eta'$ mesons as well as $B$ mesons. 
Then, the sensitivity in the large $\epsilon$ region can be improved for the long-lived scalar case. 
In the end, we would like to emphasize that the mechanism of light gauge boson production from associated U(1)-breaking scalars can be applied for a wide class of models, such as gauged U(1)$_{B-L}$ and U(1)$_{L_\alpha - L_\beta}~(\alpha,\beta=e,\mu,\tau)$ models.

\acknowledgments

We would like to thank Felix Kling and Sebastian Trojanowski for valuable comments and suggestions. 
We also thank especially Felix Kling for careful reading of the manuscript as well as providing inputs for our study. 

This work was supported by JSPS KAKENHI Grant Number JP19J13812 [KA],
~JP20K04004 [YT], 
~JP18H01210 [TA, TS],~JP18K03651, and MEXT KAKENHI Grant Number JP18H05543 [TS].

\bibliography{ref}

\providecommand{\href}[2]{#2}\begingroup\raggedright\begin{thebibliography}{10}

\bibitem{Feng:2017uoz}
J.~L. Feng, I.~Galon, F.~Kling, and S.~Trojanowski, ``{ForwArd Search
  ExpeRiment at the LHC},''
  \href{http://dx.doi.org/10.1103/PhysRevD.97.035001}{{\em Phys. Rev. D}
  {\bfseries 97} no.~3, (2018) 035001},
  \href{http://arxiv.org/abs/1708.09389}{{\ttfamily arXiv:1708.09389
  [hep-ph]}}.

\bibitem{Ariga:2018pin}
{\bfseries FASER} Collaboration, A.~Ariga {\em et al.}, ``{Technical Proposal
  for FASER: ForwArd Search ExpeRiment at the LHC},''
  \href{http://arxiv.org/abs/1812.09139}{{\ttfamily arXiv:1812.09139
  [physics.ins-det]}}.

\bibitem{Ariga:2019ufm}
{\bfseries FASER} Collaboration, A.~Ariga {\em et al.}, ``{FASER: ForwArd
  Search ExpeRiment at the LHC},''
  \href{http://arxiv.org/abs/1901.04468}{{\ttfamily arXiv:1901.04468
  [hep-ex]}}.

\bibitem{Ariga:2018uku}
{\bfseries FASER} Collaboration, A.~Ariga {\em et al.}, ``{FASER's physics
  reach for long-lived particles},''
  \href{http://dx.doi.org/10.1103/PhysRevD.99.095011}{{\em Phys. Rev. D}
  {\bfseries 99} no.~9, (2019) 095011},
  \href{http://arxiv.org/abs/1811.12522}{{\ttfamily arXiv:1811.12522
  [hep-ph]}}.

\bibitem{Feng:2017vli}
J.~L. Feng, I.~Galon, F.~Kling, and S.~Trojanowski, ``{Dark Higgs bosons at the
  ForwArd Search ExpeRiment},''
  \href{http://dx.doi.org/10.1103/PhysRevD.97.055034}{{\em Phys. Rev. D}
  {\bfseries 97} no.~5, (2018) 055034},
  \href{http://arxiv.org/abs/1710.09387}{{\ttfamily arXiv:1710.09387
  [hep-ph]}}.

\bibitem{Boiarska:2019vid}
I.~Boiarska, K.~Bondarenko, A.~Boyarsky, M.~Ovchynnikov, O.~Ruchayskiy, and
  A.~Sokolenko, ``{Light scalar production from Higgs bosons and FASER 2},''
  \href{http://dx.doi.org/10.1007/JHEP05(2020)049}{{\em JHEP} {\bfseries 05}
  (2020) 049}, \href{http://arxiv.org/abs/1908.04635}{{\ttfamily
  arXiv:1908.04635 [hep-ph]}}.

\bibitem{Feng:2018pew}
J.~L. Feng, I.~Galon, F.~Kling, and S.~Trojanowski, ``{Axionlike particles at
  FASER: The LHC as a photon beam dump},''
  \href{http://dx.doi.org/10.1103/PhysRevD.98.055021}{{\em Phys. Rev. D}
  {\bfseries 98} no.~5, (2018) 055021},
  \href{http://arxiv.org/abs/1806.02348}{{\ttfamily arXiv:1806.02348
  [hep-ph]}}.

\bibitem{Okada:2019opp}
N.~Okada and D.~Raut, ``{Hunting Inflaton at FASER},''
  \href{http://arxiv.org/abs/1910.09663}{{\ttfamily arXiv:1910.09663
  [hep-ph]}}.

\bibitem{Kling:2018wct}
F.~Kling and S.~Trojanowski, ``{Heavy Neutral Leptons at FASER},''
  \href{http://dx.doi.org/10.1103/PhysRevD.97.095016}{{\em Phys. Rev. D}
  {\bfseries 97} no.~9, (2018) 095016},
  \href{http://arxiv.org/abs/1801.08947}{{\ttfamily arXiv:1801.08947
  [hep-ph]}}.

\bibitem{Kling:2020mch}
F.~Kling and S.~Trojanowski, ``{Looking forward to test the KOTO anomaly with
  FASER},'' \href{http://dx.doi.org/10.1103/PhysRevD.102.015032}{{\em Phys.
  Rev. D} {\bfseries 102} no.~1, (2020) 015032},
  \href{http://arxiv.org/abs/2006.10630}{{\ttfamily arXiv:2006.10630
  [hep-ph]}}.

\bibitem{Okun:1982xi}
L.~Okun, ``{LIMITS OF ELECTRODYNAMICS: PARAPHOTONS?},'' {\em Sov. Phys. JETP}
  {\bfseries 56} (1982) 502.

\bibitem{Galison:1983pa}
P.~Galison and A.~Manohar, ``{TWO Z's OR NOT TWO Z's?},''
  \href{http://dx.doi.org/10.1016/0370-2693(84)91161-4}{{\em Phys. Lett. B}
  {\bfseries 136} (1984) 279--283}.

\bibitem{Holdom:1985ag}
B.~Holdom, ``{Two U(1)'s and Epsilon Charge Shifts},''
\href{http://dx.doi.org/10.1016/0370-2693(86)91377-8}{{\em Phys. Lett.}
  {\bfseries B166} (1986) 196--198}.

\bibitem{Foot:1991kb}
R.~Foot and X.-G. He, ``{Comment on Z Z-prime mixing in extended gauge
  theories},''
\href{http://dx.doi.org/10.1016/0370-2693(91)90901-2}{{\em Phys. Lett.}
  {\bfseries B267} (1991) 509--512}.

\bibitem{Babu:1997st}
K.~S. Babu, C.~F. Kolda, and J.~March-Russell, ``{Implications of generalized Z
  - Z-prime mixing},'' \href{http://dx.doi.org/10.1103/PhysRevD.57.6788}{{\em
  Phys. Rev.} {\bfseries D57} (1998) 6788--6792},
\href{http://arxiv.org/abs/hep-ph/9710441}{{\ttfamily arXiv:hep-ph/9710441
  [hep-ph]}}.

\bibitem{Boehm:2003hm}
C.~Boehm and P.~Fayet, ``{Scalar dark matter candidates},''
  \href{http://dx.doi.org/10.1016/j.nuclphysb.2004.01.015}{{\em Nucl. Phys. B}
  {\bfseries 683} (2004) 219--263},
  \href{http://arxiv.org/abs/hep-ph/0305261}{{\ttfamily arXiv:hep-ph/0305261}}.

\bibitem{Pospelov:2008zw}
M.~Pospelov, ``{Secluded U(1) below the weak scale},''
  \href{http://dx.doi.org/10.1103/PhysRevD.80.095002}{{\em Phys. Rev. D}
  {\bfseries 80} (2009) 095002},
  \href{http://arxiv.org/abs/0811.1030}{{\ttfamily arXiv:0811.1030 [hep-ph]}}.

\bibitem{Fabbrichesi:2020wbt}
M.~Fabbrichesi, E.~Gabrielli, and G.~Lanfranchi, ``{The Dark Photon},''
  \href{http://arxiv.org/abs/2005.01515}{{\ttfamily arXiv:2005.01515
  [hep-ph]}}.

\bibitem{Jodlowski:2019ycu}
K.~Jod\l~owski, F.~Kling, L.~Roszkowski, and S.~Trojanowski, ``{Extending the
  reach of FASER, MATHUSLA, and SHiP towards smaller lifetimes using secondary
  particle production},''
  \href{http://dx.doi.org/10.1103/PhysRevD.101.095020}{{\em Phys. Rev. D}
  {\bfseries 101} no.~9, (2020) 095020},
  \href{http://arxiv.org/abs/1911.11346}{{\ttfamily arXiv:1911.11346
  [hep-ph]}}.

\bibitem{Aaij:2015tna}
{\bfseries LHCb} Collaboration, R.~Aaij {\em et al.}, ``{Search for
  hidden-sector bosons in $B^0 \!\to K^{*0}\mu^+\mu^-$ decays},''
  \href{http://dx.doi.org/10.1103/PhysRevLett.115.161802}{{\em Phys. Rev.
  Lett.} {\bfseries 115} no.~16, (2015) 161802},
  \href{http://arxiv.org/abs/1508.04094}{{\ttfamily arXiv:1508.04094
  [hep-ex]}}.

\bibitem{Aaij:2016qsm}
{\bfseries LHCb} Collaboration, R.~Aaij {\em et al.}, ``{Search for long-lived
  scalar particles in $B^+ \to K^+ \chi (\mu^+\mu^-)$ decays},''
  \href{http://dx.doi.org/10.1103/PhysRevD.95.071101}{{\em Phys. Rev. D}
  {\bfseries 95} no.~7, (2017) 071101},
  \href{http://arxiv.org/abs/1612.07818}{{\ttfamily arXiv:1612.07818
  [hep-ex]}}.

\bibitem{Bezrukov:2013fca}
F.~Bezrukov and D.~Gorbunov, ``{Light inflaton after LHC8 and WMAP9 results},''
  \href{http://dx.doi.org/10.1007/JHEP07(2013)140}{{\em JHEP} {\bfseries 07}
  (2013) 140}, \href{http://arxiv.org/abs/1303.4395}{{\ttfamily arXiv:1303.4395
  [hep-ph]}}.

\bibitem{Winkler:2018qyg}
M.~W. Winkler, ``{Decay and detection of a light scalar boson mixing with the
  Higgs boson},'' \href{http://dx.doi.org/10.1103/PhysRevD.99.015018}{{\em
  Phys. Rev. D} {\bfseries 99} no.~1, (2019) 015018},
  \href{http://arxiv.org/abs/1809.01876}{{\ttfamily arXiv:1809.01876
  [hep-ph]}}.

\bibitem{Buschmann:2015awa}
M.~Buschmann, J.~Kopp, J.~Liu, and P.~A.~N. Machado, ``{Lepton Jets from
  Radiating Dark Matter},''
  \href{http://dx.doi.org/10.1007/JHEP07(2015)045}{{\em JHEP} {\bfseries 07}
  (2015) 045}, \href{http://arxiv.org/abs/1505.07459}{{\ttfamily
  arXiv:1505.07459 [hep-ph]}}.

\bibitem{Bauer:2018onh}
M.~Bauer, P.~Foldenauer, and J.~Jaeckel, ``{Hunting All the Hidden Photons},''
  \href{http://dx.doi.org/10.1007/JHEP07(2018)094}{{\em JHEP} {\bfseries 18}
  (2020) 094}, \href{http://arxiv.org/abs/1803.05466}{{\ttfamily
  arXiv:1803.05466 [hep-ph]}}.

\bibitem{Campabadal:2005cn}
F.~Campabadal {\em et al.}, ``{Beam tests of ATLAS SCT silicon strip detector
  modules},'' \href{http://dx.doi.org/10.1016/j.nima.2004.08.133}{{\em Nucl.
  Instrum. Meth. A} {\bfseries 538} (2005) 384--407}.

\bibitem{Ariga:2018zuc}
{\bfseries FASER} Collaboration, A.~Ariga {\em et al.}, ``{Letter of Intent for
  FASER: ForwArd Search ExpeRiment at the LHC},''
  \href{http://arxiv.org/abs/1811.10243}{{\ttfamily arXiv:1811.10243
  [physics.ins-det]}}.

\bibitem{Kling2019}
W.~F.~K. and. Private communication.

\bibitem{Pierog_2015}
T.~Pierog, I.~Karpenko, J.~M. Katzy, E.~Yatsenko, and K.~Werner, ``EPOS LHC:
  Test of collective hadronization with data measured at the CERN Large Hadron
  Collider,'' \href{http://dx.doi.org/10.1103/physrevc.92.034906}{{\em Physical
  Review C} {\bfseries 92} no.~3, (Sep, 2015) }.
  \url{http://dx.doi.org/10.1103/PhysRevC.92.034906}.

\bibitem{CRMC}
C.~Baus, T.~Pierog, and R.~Ulrich, ``Cosmic Ray Monte Carlo (CRMC),''.
  \url{https://web.ikp.kit.edu/rulrich/crmc.html}.

\bibitem{Sj_strand_2006}
T.~Sjöstrand, S.~Mrenna, and P.~Skands, ``PYTHIA 6.4 physics and manual,''
  \href{http://dx.doi.org/10.1088/1126-6708/2006/05/026}{{\em Journal of High
  Energy Physics} {\bfseries 2006} no.~05, (May, 2006) 026–026}.
  \url{http://dx.doi.org/10.1088/1126-6708/2006/05/026}.

\bibitem{Sj_strand_2008}
T.~Sjöstrand, S.~Mrenna, and P.~Skands, ``A brief introduction to PYTHIA
  8.1,'' \href{http://dx.doi.org/10.1016/j.cpc.2008.01.036}{{\em Computer
  Physics Communications} {\bfseries 178} no.~11, (Jun, 2008) 852–867}.
  \url{http://dx.doi.org/10.1016/j.cpc.2008.01.036}.

\bibitem{Skands_2014}
P.~Skands, S.~Carrazza, and J.~Rojo, ``Tuning PYTHIA 8.1: the Monash 2013
  tune,'' \href{http://dx.doi.org/10.1140/epjc/s10052-014-3024-y}{{\em The
  European Physical Journal C} {\bfseries 74} no.~8, (Aug, 2014) }.
  \url{http://dx.doi.org/10.1140/epjc/s10052-014-3024-y}.

\bibitem{Chang:2016ntp}
J.~H. Chang, R.~Essig, and S.~D. McDermott, ``{Revisiting Supernova 1987A
  Constraints on Dark Photons},''
  \href{http://dx.doi.org/10.1007/JHEP01(2017)107}{{\em JHEP} {\bfseries 01}
  (2017) 107}, \href{http://arxiv.org/abs/1611.03864}{{\ttfamily
  arXiv:1611.03864 [hep-ph]}}.

\bibitem{Kling-Trojanowski2020}
F.~Kling and S.~Trojanowski. Private communication and discussion.

\bibitem{Lees:2014xha}
{\bfseries BaBar} Collaboration, J.~Lees {\em et al.}, ``{Search for a Dark
  Photon in $e^+e^-$ Collisions at BaBar},''
  \href{http://dx.doi.org/10.1103/PhysRevLett.113.201801}{{\em Phys. Rev.
  Lett.} {\bfseries 113} no.~20, (2014) 201801},
  \href{http://arxiv.org/abs/1406.2980}{{\ttfamily arXiv:1406.2980 [hep-ex]}}.

\bibitem{Batley:2015lha}
{\bfseries NA48/2} Collaboration, J.~Batley {\em et al.}, ``{Search for the
  dark photon in $\pi^0$ decays},''
  \href{http://dx.doi.org/10.1016/j.physletb.2015.04.068}{{\em Phys. Lett. B}
  {\bfseries 746} (2015) 178--185},
  \href{http://arxiv.org/abs/1504.00607}{{\ttfamily arXiv:1504.00607
  [hep-ex]}}.

\bibitem{Bernhard:2020vca}
{\bfseries NA64, Physics Beyond Collider Conventional Beams working group}
  Collaboration, J.~Bernhard, ``{Status and Plans for the NA64 Experiment},''
  \href{http://dx.doi.org/10.1088/1742-6596/1468/1/012023}{{\em J. Phys. Conf.
  Ser.} {\bfseries 1468} no.~1, (2020) 012023}.

\bibitem{Anastasi:2016ktq}
{\bfseries KLOE-2} Collaboration, A.~Anastasi {\em et al.}, ``{Limit on the
  production of a new vector boson in $\mathrm{e^+ e^-}\rightarrow {\rm
  U}\gamma$, U$\rightarrow \pi^+\pi^-$ with the KLOE experiment},''
  \href{http://dx.doi.org/10.1016/j.physletb.2016.04.019}{{\em Phys. Lett. B}
  {\bfseries 757} (2016) 356--361},
  \href{http://arxiv.org/abs/1603.06086}{{\ttfamily arXiv:1603.06086
  [hep-ex]}}.

\bibitem{Aaij:2019bvg}
{\bfseries LHCb} Collaboration, R.~Aaij {\em et al.}, ``{Search for
  $A'\to\mu^+\mu^-$ Decays},''
  \href{http://dx.doi.org/10.1103/PhysRevLett.124.041801}{{\em Phys. Rev.
  Lett.} {\bfseries 124} no.~4, (2020) 041801},
  \href{http://arxiv.org/abs/1910.06926}{{\ttfamily arXiv:1910.06926
  [hep-ex]}}.

\bibitem{Riordan:1987aw}
E.~Riordan {\em et al.}, ``{A Search for Short Lived Axions in an Electron Beam
  Dump Experiment},'' \href{http://dx.doi.org/10.1103/PhysRevLett.59.755}{{\em
  Phys. Rev. Lett.} {\bfseries 59} (1987) 755}.

\bibitem{Marsicano:2018krp}
L.~Marsicano, M.~Battaglieri, M.~Bondi', C.~R. Carvajal, A.~Celentano,
  M.~De~Napoli, R.~De~Vita, E.~Nardi, M.~Raggi, and P.~Valente, ``{Dark photon
  production through positron annihilation in beam-dump experiments},''
  \href{http://dx.doi.org/10.1103/PhysRevD.98.015031}{{\em Phys. Rev. D}
  {\bfseries 98} no.~1, (2018) 015031},
  \href{http://arxiv.org/abs/1802.03794}{{\ttfamily arXiv:1802.03794
  [hep-ex]}}.

\bibitem{Blumlein:2013cua}
J.~Bl{\"{u}}umlein and J.~Brunner, ``{New Exclusion Limits on Dark Gauge Forces
  from Proton Bremsstrahlung in Beam-Dump Data},''
  \href{http://dx.doi.org/10.1016/j.physletb.2014.02.029}{{\em Phys. Lett. B}
  {\bfseries 731} (2014) 320--326},
  \href{http://arxiv.org/abs/1311.3870}{{\ttfamily arXiv:1311.3870 [hep-ph]}}.

\end{thebibliography}\endgroup

\end{document}